\documentclass[aip,reprint]{revtex4-1}

\usepackage{graphicx}% Include figure files
\usepackage{amsmath}
\usepackage{comment}

\draft % marks overfull lines with a black rule on the right

\begin{document}

%\preprint{}

\title{Lasing from a Quantum-Dot-Like Buried Heterostructure in an InP Nanobeam Cavity} %Title of paper



\author{Valdemar Bille-Lauridsen}
\email[]{vchbi@dtu.dk}
\affiliation{Department of Electrical and Photonics Engineering, Technical University of Denmark, Building 343, 2800 Kongens Lyngby, Denmark}
\affiliation{NanoPhoton - Center for Nanophotonics, Technical University of Denmark, Building 343, 2800 Kongens Lyngby, Denmark}

%\author{}
%\thanks{}
%\affiliation{}

\author{Rasmus Jarbøl}
\affiliation{Department of Electrical and Photonics Engineering, Technical University of Denmark, Building 343, 2800 Kongens Lyngby, Denmark}

\author{Meng Xiong}
\affiliation{Department of Electrical and Photonics Engineering, Technical University of Denmark, Building 343, 2800 Kongens Lyngby, Denmark}
\affiliation{NanoPhoton - Center for Nanophotonics, Technical University of Denmark, Building 343, 2800 Kongens Lyngby, Denmark}

\author{Aurimas Sakanas}
\affiliation{Department of Electrical and Photonics Engineering, Technical University of Denmark, Building 343, 2800 Kongens Lyngby, Denmark}

\author{Elizaveta Semenova}
\affiliation{Department of Electrical and Photonics Engineering, Technical University of Denmark, Building 343, 2800 Kongens Lyngby, Denmark}
\affiliation{NanoPhoton - Center for Nanophotonics, Technical University of Denmark, Building 343, 2800 Kongens Lyngby, Denmark}

\author{Kresten Yvind}
\affiliation{Department of Electrical and Photonics Engineering, Technical University of Denmark, Building 343, 2800 Kongens Lyngby, Denmark}
\affiliation{NanoPhoton - Center for Nanophotonics, Technical University of Denmark, Building 343, 2800 Kongens Lyngby, Denmark}

\author{Jesper Mørk}
\affiliation{Department of Electrical and Photonics Engineering, Technical University of Denmark, Building 343, 2800 Kongens Lyngby, Denmark}
\affiliation{NanoPhoton - Center for Nanophotonics, Technical University of Denmark, Building 343, 2800 Kongens Lyngby, Denmark}


%\date{}

\begin{abstract}
We report lasing from a lithographically defined buried heterostructure with an estimated lateral footprint of $(107\,\mathrm{nm})^2$, embedded in an InP photonic-crystal nanobeam cavity. 
This represents the smallest laterally confined buried heterostructure gain region from which lasing has been observed. 
Despite etching of the active region during cavity definition and the associated risk of surface-related nonradiative recombination, optically pumped devices exhibit a clear lasing threshold and a narrow linewidth. 
By systematically varying the buried heterostructure size, we investigate how the lasing threshold depends on the active volume under optical pumping. The estimated intrinsic threshold under ideal carrier injection is 57 nW, comparable to values reported for single quantum-dot nanolasers, highlighting the potential of quantum-dot-scale buried heterostructures as deterministic, scalable gain media for nanophotonic lasers.

% Journal reference + DOI for arXiv version
\medskip
\noindent\textit{Published in APL Photonics \textbf{11}, 071301 (2026). DOI: 10.1063/5.0334223.}
\end{abstract}


\maketitle %\maketitle must follow title, authors, abstract and \pacs

%\section{Introduction}

Ultracompact semiconductor lasers are a key enabling technology for densely integrated photonic circuits. In these lasers, low energy consumption, high optical confinement, and precise spatial definition of the gain region are key requirements \cite{obrienPhotonicQuantumTechnologies2009, notomiManipulatingLightStrongly2010,jeongElectricallyDrivenNanobeam2013, dimopoulosExperimentalDemonstrationNanolaser2023}.
Photonic crystal nanocavities provide small mode volumes and high quality factors \cite{gongNanobeamPhotonicCrystal2010}, enabling strong light–matter interaction \cite{ohtaStrongCouplingPhotonic2011}. As cavities and active regions decrease in size, it becomes increasingly challenging to achieve a large controlled overlap between the optical cavity field and the gain region.
Self-assembled quantum dots (QDs) offer excellent emission properties and have enabled record-low-threshold nanolasers \cite{otaThresholdlessQuantumDot2017}. Yet their stochastic nucleation leads to poor spatial alignment with cavity modes, lowering fabrication yield, and limiting scalability \cite{sapienzaNanoscaleOpticalPositioning2015}. Deterministic alternatives based on micropillars, nanorods or site-selected growth improve positioning, but these approaches have different challenges~\cite{Schneider2009,Poole_2010,Haffouz2018,holewaOpticalPropertiesSiteSelectively2021,holewaSolidstateSinglephotonSources2025}. In particular, these approaches do not readily translate to planar cavities such as two-dimensional photonic crystals or nanobeams with small mode volumes, where efficient operation requires the active region to be precisely located at the cavity field maximum. %Find source
This is especially true for bowtie cavities, where the central features can reach 10 nm \cite{xiongNanolaserExtremeDielectric2025}. %more sources with example devices

%cite alisha?
%Other design trade offs... small gain/dipole moment, carrier losses etc. [SOURCEs].

Buried heterostructures (BHs), which are laterally confined quantum wells formed by lithography and regrowth, provide a scalable solution offering strong lateral carrier confinement and precise spatial definition of the active region \cite{matsuoHighspeedUltracompactBuried2010,SakanasBHLaserPhD,dimopoulosElectricallyDrivenPhotonicCrystal2022}. Recently, photonic crystal lasers have achieved exceptionally low threshold using such a BH structure \cite{matsuoUltralowOperatingEnergy2013,dimopoulosExperimentalDemonstrationNanolaser2023}. However, in previous demonstrations, the BH dimensions are several hundred nanometers, far from the quantum-dot regime.

\begin{figure}
    \centering
    \includegraphics[width=0.95\linewidth]{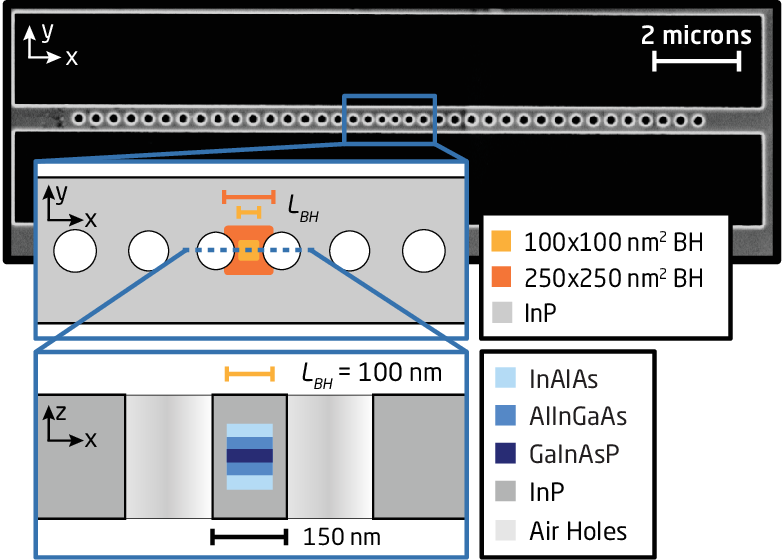}
    \caption{Scanning electron microscope image of a fabricated nanobeam laser. A sketch of the central region is shown below, highlighting the scale of two different BH lengths $l_\mathrm{BH}$ as compared to the holes. At the bottom, the cross-section of the BH is illustrated. }
    \label{fig:device}
\end{figure}

In this work, we experimentally demonstrate lasing from buried heterostructures with estimated lateral dimensions between $\sim(107\, \mathrm{nm})^2$ and $(216 \, \mathrm{nm})^2$ embedded in an InP photonic crystal nanobeam cavity. This represents the smallest laterally confined BH active region from which lasing has been reported. Despite the nearby etched surfaces, where nonradiative surface recombination is typically dominant in nanoscale III–V emitters \cite{wangOpticalPropertiesSingle2004,manna_surface_2020}, we observe a clear threshold transition and narrow linewidth under optical pumping. Lasing is enabled by careful surface passivation \cite{berdnikovEfficientPassivationIIIAsP2025} and strong spatial overlap between the cavity mode and the nanoscale BH. 
The smallest BH structures in this work are comparable to the size of early telecom-wavelength quantum dots, which were reported with base diameters exceeding 100 nm \cite{anantathanasarnWavelengthtunable155mmRegion2005}.

\begin{figure}
    \centering
    \includegraphics[width=1\linewidth]{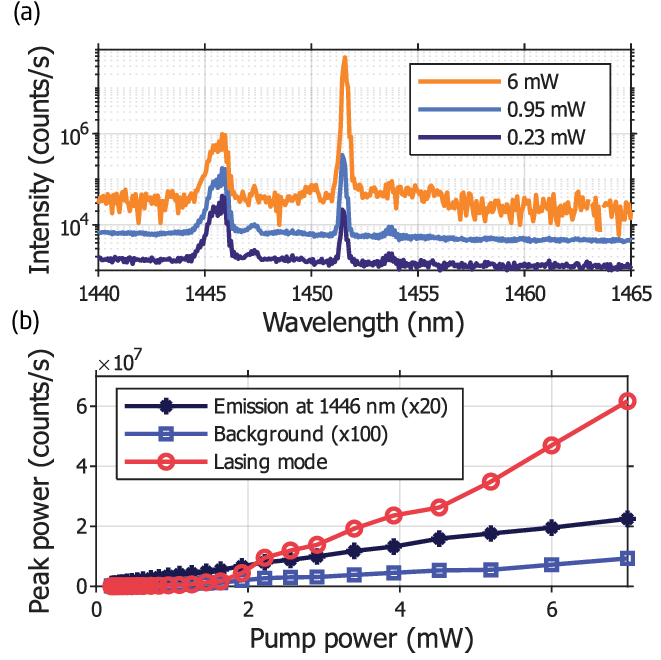}
    \caption{(a) Emission spectra around the lasing wavelength from the laser with $l_\mathrm{BH}=107$ nm as measured far below, near, and above threshold. The spectrometer's exposure time is greatly reduced above the threshold, leading to increased noise.
    (b) Pump power dependent peak emission intensity of the lasing mode compared to the emission around $1446\, \mathrm{nm}$ and the background, which is taken at $1460\, \mathrm{nm}$.}
    \label{fig:lasing}
\end{figure}

%\section{Device Fabrication}

The active region is based on an InP membrane incorporating a single GaInAsP quantum well and barrier layers [Fig. \ref{fig:device}] grown by metal-organic vapour-phase epitaxy. The buried heterostructure is defined by electron-beam lithography and dry etching of a nanoscale mesa, followed by regrowth to laterally encapsulate the active material. The BH fabrication method is the same as described in Ref. [\onlinecite{dimopoulosElectricallyDrivenPhotonicCrystal2022}]. 

Scanning electron microscope (SEM) inspection of the Hydrogen Silsesquioxane (HSQ) mask, used for defining the BH structures, reveals that the realized mask sizes are systematically smaller than the nominal design values. Buried heterostructures with side lengths $l_{BH}=84(\pm5)$~nm up to $l_{BH}=212(\pm6)$~nm were fabricated, assuming that the active material size corresponds to the HSQ mask size.

Following BH fabrication, the photonic crystal nanobeam cavities are subsequently defined by a separate lithography and etching process. The photonic crystal nanobeam cavities consist of a one-dimensional array of air holes with a linear tapered radius profile \cite{gongNanobeamPhotonicCrystal2010, quanDeterministicDesignWavelength2011}. The cavity fabrication steps are also described in Ref. [\onlinecite{dimopoulosElectricallyDrivenPhotonicCrystal2022}].

The cavity designs exhibit simulated quality factors exceeding $Q\simeq 10^5$ and effective mode volumes of $V_m \simeq 5(\lambda/2n)^3$. The fabricated devices have quality factors of  $Q = 24.0(\pm1.6)\times10^3$, as determined by measuring nominally identical passive cavities following the approach of Galli et al.~\cite{galliLightScatteringFano2009}. The reduction in quality factor is attributed primarily to fabrication-induced surface roughness, consistent with previous reports on InP nanobeam cavities~\cite{ohtaStrongCouplingPhotonic2011}.

The separation between the two central holes in the cavity design is 150 nm. Therefore, BHs with lateral dimensions exceeding this separation overlap with the etched air holes and become partially etched through during cavity definition, resulting in an exposed air–BH interface when $l_{BH}>150$ nm [Fig.~\ref{fig:device}]. For smaller BHs, etch-through may still occur due to lithographic misalignment between the BH and cavity. 
Since surface-related nonradiative recombination can be significant in exposed III–V nanostructures~\cite{higuera-rodriguezUltralowSurfaceRecombination2017}, surface passivation is applied after cavity definition and membranization. The native oxide is removed using $\mathrm{NH_4OH}$ and $\mathrm{(NH_4)_2S}$, after which a 7~nm layer of Al$_2$O$_3$ is deposited by atomic layer deposition for surface passivation. This passivation process covers the exposed nanobeam surfaces, including any exposed BH interfaces. The steps of the surface passivation method are detailed in Ref. [\onlinecite{berdnikovEfficientPassivationIIIAsP2025}].

We now present the performance of a device with an active region length of $l_{\mathrm{BH}}=107 (\pm 5)~\mathrm{nm}$. This was the device with the smallest BH size for which lasing was confirmed. The devices are characterized using two different experimental setups. In setup 1, the device is studied at room temperature under continuous-wave optical pumping at $\lambda_p=980\,\mathrm{nm}$ through an objective with numerical aperture of 0.65, giving a diffraction-limited spot size of  $A_\mathrm{spot}\sim(668 \,\mathrm{nm})^2$. The output is collected using a spectrometer (SR-500i, Andor) with an InGaAs detector (iDus 1.7 \textmu m, Andor), enabling reliable detection of low output powers.

Figure~\ref{fig:lasing} shows the emission spectra of the device. The dominant lasing peak is observed at $1452\,\mathrm{nm}$. An additional peak is observed at $1446\,\mathrm{nm}$ together with a broadband background signal. Control measurements confirm that these features do not originate from the nanobeam device or the nanoscale BH, but from a much larger active region located nearby. When the pump spot is moved away from the nanobeam, the lasing mode disappears, whereas the $1446\,\mathrm{nm}$ peak and background emission remain \cite{ourDATA}.
As the pump power increases, the mode at $1452\,\mathrm{nm}$ becomes dominant and exhibits a pronounced nonlinear increase in intensity, indicating the onset of lasing at an incident pump power just below $2\,\mathrm{mW}$, as shown in Fig.~\ref{fig:lasing}(b). 

%The two small mode visible below threshold show sign until it clamps at around 1~mW, confirming that stimulated emission occurs only for the lasing mode above a pump power of approximately 1~mW. %Svært at vise clamping ikke nok data points, hvad med baggrund noise, osv.

\begin{figure}
    \centering
    \includegraphics[width=1\linewidth]{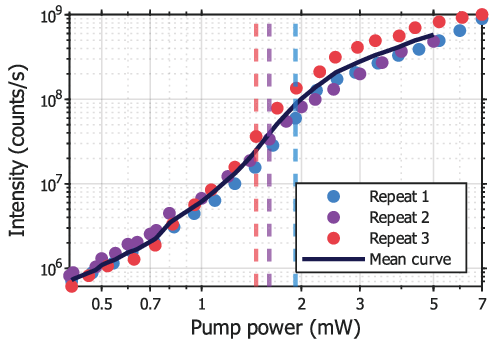}
    \caption{Output versus incident pump power for the laser with $l_\mathrm{BH}=107$ nm. Three repeated measurements, the average curve, and the extracted thresholds (vertical dashed lines) are shown.}
    \label{fig:LL-repeated}
\end{figure}

Measured input-output characteristics of the same laser are shown in Fig.~\ref{fig:LL-repeated} on a log–log scale. A characteristic S-shaped input–output curve is observed. From this input-output characteristic, the threshold is extracted as the pump power at which the derivative of the curve reaches its maximum \cite{saldutti_onset_2024}. To account for variations in in-coupling efficiency due to pump spot alignment, the measurement was repeated three times. Although the pump spot is substantially larger than the BH dimensions, the injection efficiency is very sensitive to small changes in the pump spot position. The extracted thresholds from each trial are shown in the figure, indicating significant variation between measurements with a mean threshold at 1.6 mW and a standard deviation of 0.24 mW. With diffraction-limiteded pump spot, a pump power of $1.6\,\mathrm{mW}$ corresponds to an incident optical pump density of $3.6\times10^5 ~\mathrm{W/cm^2}$. We emphasize that only a very small fraction of the incident pump power gets absorbed leading to electron excitation.

\begin{figure}
    \centering
    \includegraphics[width=0.95\linewidth]{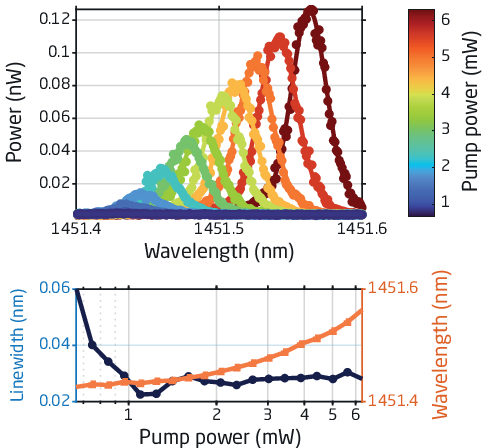}
    \caption{(a) Emission spectra of laser with $l_\mathrm{BH}=107$ nm as measured in setup 2. Solid curves are Gaussian fits to the spectra. (b) From the fit of the spectra, peak wavelengths and widths are extracted and plotted as a function of the pump power. The pump powers in this figure cannot be directly compared to those in Figs. \ref{fig:lasing}-\ref{fig:LL-repeated}, as the injection efficiencies between the setups will differ.}
    \label{fig:Spectra}
\end{figure}

To further confirm that the lasing threshold is reached, the devices were also characterized in a secound setup. In setup 2 the measurements were likewise performed at room temperature using the same continuous-wave pump, but with an optical spectrum analyzer (AQ6370D, Yokogawa), providing improved spectral resolution. The results are shown in Fig.~\ref{fig:Spectra}. Here, it is seen that the extracted linewidths are near the resolution limit of 0.02~nm, and are therefore likely instrument limited, consistent with lasing operation. The linewidth extracted at the lowest pump powers is subject to significant uncertainty due to the limited signal-to-noise ratio. The redshift at the highest pump powers is consistent with pump-induced heating and associated refractive-index changes \cite{nomuraRoomTemperatureContinuouswave2006, lingluGainCompressionThermal2009}.
%In this figure, all data points are taken above threshold, as the noise floor was too high to measure the signal below threshold [How do we know, justify].
%Repeat this measurement as well to get statistics

%\section{Scaling and gain}

\begin{figure}
    \centering
    \includegraphics[width=0.95\linewidth]{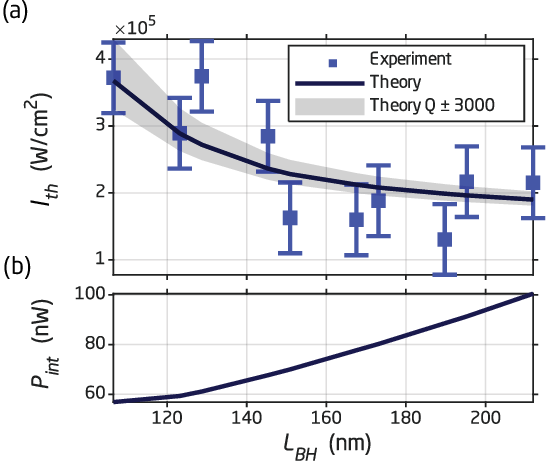}
    \caption{(a) Threshold value of incident pump power density $I_\mathrm{th}$ versus BH length. The error bars indicate the standard deviation as found in Fig. \ref{fig:LL-repeated}. The solid line is a fit to the threshold expression, Eq.~\eqref{eq:density_threshold}. (b) Corresponding estimated value of pump power using  Eq.~\eqref{eq:Threshold_power}  with an ideal injection efficiency $\eta_\mathrm{abs}=1$.}
    \label{fig:Thresholds_vs_theory}
\end{figure}

We next consider the comparative performance of lasers with larger BH sizes.
In addition to providing deterministic placement of a small active region, the lithographic definition of the BH also enables systematic control over its size. This allows us to study how the threshold of the nanolaser devices depends on the size of the active medium. 
An input-output curve is captured for all fabricated devices with BH sizes ranging from 107 to 212 nm. For all these devices, a clear threshold was found using the same method as described previously. The extracted thresholds are shown as a function of the BH size in figure \ref{fig:Thresholds_vs_theory}(a).

The lasing threshold can be estimated from the steady-state gain-loss condition by assuming a logarithmic material gain,
$g(N)=g_0\ln(N/N_\mathrm{tr})$, where $g_0$ is the gain coefficient and $N_\mathrm{tr}$ is the transparency carrier density \cite{coldren2012diode}.
At threshold, the modal gain equals the cavity loss, $\Gamma v_g g(N_\mathrm{th}) = 1/\tau_p$, where $\Gamma$ is the optical confinement factor,
$v_g$ is the group velocity in the cavity, and $\tau_p$ is the photon lifetime. This yields the threshold carrier density
\begin{equation}
N_\mathrm{th} = N_\mathrm{tr}\exp\!\left(\frac{1}{\Gamma v_g g_0 \tau_p}\right).
\label{eq:N_th}
\end{equation}
From this expression, it is seen that a laser will operate close to the transparency $N_\mathrm{th}\sim N_\mathrm{tr}$, when the gain and confinement factor are high compared to the loss. 
The corresponding pump-power threshold can be estimated as
\begin{equation}
P_\mathrm{th}
=
\frac{h\nu_p}{\eta_{\mathrm{abs}}}\,
\frac{V_{\mathrm{act}}}{\tau_{\mathrm{eff}}}\,
N_\mathrm{th},
\label{eq:Threshold_power}
\end{equation}
where $h\nu_p$ is the pump-photon energy, $V_\mathrm{act}$ is the active-region volume, $\eta_{\rm abs}$ is the injection efficiency, and
$\tau_\mathrm{eff}$ is the effective carrier lifetime defined by $\tau_\mathrm{eff}^{-1}=\tau_r^{-1}+\tau_{nr}^{-1}$
with radiative and nonradiative lifetimes $\tau_r$ and $\tau_{nr}$, respectively.

To better understand the measured results and the potential of the devices, we calculate an intrinsic threshold, which we define as the threshold under ideal carrier injection. In the case of electrically pumped devices, $\eta_\mathrm{abs}$ can be close to unity, although leakage paths may play a role \cite{marchalCarrierTransportElectricallyDriven2025}. Therefore, we define the intrinsic threshold $P_{\rm int}$ as the threshold power assuming perfect injection $\eta_\mathrm{abs}=1$.
In this case, smaller active regions are expected to yield lower thresholds \cite{dimopoulosExperimentalDemonstrationNanolaser2023}. This follows directly from Eq.~\eqref{eq:Threshold_power}, as $P_{\mathrm{th}}$ scales approximately with $V_\mathrm{act}/\eta_{\mathrm{abs}}$. 

However, in these experiments, the devices are optically pumped with a spot that is much larger than the active region. Consequently, only a small fraction of the incident power overlaps the BH. We account for this by expressing the absorbed pump fraction as
\begin{equation}
\eta_{\mathrm{abs}} = \eta_0\frac{A_{\mathrm{act}}}{A_{\mathrm{spot}}},
\label{eq:pumpspot_injection_eff}
\end{equation}
where $\eta_0$ is the absorption efficiency for perfect spatial overlap, $A_{\mathrm{spot}}$ is the pump-spot area, and $A_{\mathrm{act}}$ is the lateral area of the active region. The active-layer thickness is $t = 8\,\mathrm{nm}$ for all BH devices. In terms of power density, $I_\mathrm{th}=P_{\rm th}/A_\mathrm{spot}$ the measured threshold becomes
\begin{equation}
I_\mathrm{th}
=
\frac{h\nu_pt}{\eta_0\tau_{\mathrm{eff}}}\,
N_\mathrm{tr}\exp\!\left(\frac{1}{\Gamma v_g g_0 \tau_p}\right),
\label{eq:density_threshold}
\end{equation}
which is now not explicitly dependent on $A_\mathrm{\rm act}$,  in this spot-limited regime.

To estimate the threshold scaling and intrinsic thresholds, we fit steady-state semiconductor laser rate equations \cite{coldren2012diode} to the experimental data. 
All model parameters were fixed except the spontaneous emission factor $\beta$, the absorption efficiency prefactor $\eta_0$, and the output coupling efficiency $\eta_\mathrm{out}$. 
Among these, only $\eta_0$ affects the extracted threshold shown in Fig.~\ref{fig:Thresholds_vs_theory}. 
The remaining parameters are taken from literature values and are summarized in Table~\ref{tab:rate_equation_parameters}. 
In the calculations, only the active volume $V_\mathrm{act}$ and the optical confinement factor $\Gamma$ were varied between devices. The confinement factors were calculated from the simulated cavity-mode profile and the estimated BH sizes, ranging from $\Gamma=0.0018$ for $l_{BH}=107~\mathrm{nm}$ to $\Gamma=0.0046$ for $l_{BH}=212~\mathrm{nm}$.

\begin{table}
\caption{Parameters used in the steady-state rate-equation model.
The fitted absorbed pump fraction $\eta_\mathrm{0}$, is one order of magnitude below the theoretical upper limit of $17\times10^{-3}$ as found by assuming that the absorption coefficient is $\alpha=3\times10^4 \,\mathrm{cm^{-1}}$,  similar to what is measured for InGaAs at a pump wavelength of 980~nm~\cite{InGaAs_abs}, and that the reflection from the top of InP surface is $27\%$.
}
\label{tab:rate_equation_parameters}
\centering
\begin{tabular}{l l c}
\hline\hline
Symbol & Description & Value \\ 
\hline
%$Q$ 
%& Cavity quality factor 
%& $2.4\times10^{4}$ \\

%g0 i Coldren er 1800cm^-1 for InGaAs 60Å WQ og Ntr 2.2*10^-18 cm^-3
%med gd=g0/Ntr giver det gd=8*10^-16 cm^2

%Fra Kristian Ntr 1.5*10^-18, 
$g_0$ 
& Material gain parameter \cite{Seegert_PhD}
& $3000\ \mathrm{cm^{-1}}$ \\

$N_\mathrm{tr}$ 
& Transparency carrier density \cite{Seegert_PhD}
& $1.5\times10^{18}\ \mathrm{cm^{-3}}$ \\

$\tau_r$ 
& Radiative carrier lifetime \cite{matsuoUltralowOperatingEnergy2013}
& $2\ \mathrm{ns}$  \\

$\tau_{nr}$ 
& Nonradiative carrier lifetime \cite{berdnikovEfficientPassivationIIIAsP2025}
& $5\ \mathrm{ns}$  \\

$\beta$ 
& Spontaneous emission factor
& $0.03$ \\

%$n_g$ 
%& Group index 
%& $3.2$ \\

%$\lambda_l$ 
%& Lasing wavelength 
%& $1.45\ \mu\mathrm{m}$ \\

%$\lambda_p$ 
%%& Pump wavelength 
%& $980\ \mathrm{nm}$ \\

$\eta_\mathrm{0}$
& Absorbed pump fraction
& $1.35\times10^{-3}$ \\

$\eta_\mathrm{out}$ 
& Output coupling efficiency 
& $0.6\times10^{-3}$ \\

%$t_\mathrm{act}$ 
%& Active region thickness 
%& $8.1\ \mathrm{nm}$ \\
\hline\hline
\end{tabular}
\end{table}

This model captures the overall trend observed in Fig.~\ref{fig:Thresholds_vs_theory}(a). 
The remaining deviation between experiment and theory is attributed to device-to-device variations that are not included in the model. 
In particular, the almost identical pair of devices with $l_{\mathrm{BH}}=145$~nm, and $l_{\mathrm{BH}}=151$~nm, exhibit noticeably different thresholds despite only a 6~nm difference in the BH size. %Such variations are expected given the sensitivity of nanobeam lasers to small changes in cavity quality factor $Q$, optical confinement factor $\Gamma$, and effective nonradiative lifetime $\tau_{nr}$, all of which can vary due to fabrication imperfections and regrowth nonuniformities.

The scaling of the threshold with active region size depends on the carrier injection scheme. 
Under an ideal injection scheme, such as electrical injection, the relevant quantity is the total number of carriers required to reach transparency, $V_{\mathrm{act}}N_\mathrm{tr}$. Reducing the BH size therefore lowers the intrinsic threshold power $P_\mathrm{int}$, as seen in Fig. \ref{fig:Thresholds_vs_theory}(b).
In this regime, the increase in threshold carrier density associated with a reduced optical confinement factor [Eq.~\eqref{eq:N_th}] leads to the flattening of the intrinsic threshold $P_\mathrm{int}$ at the smallest BH sizes.
In contrast, under optical pumping with a fixed pump spot, the measured threshold density $I_\mathrm{th}$ increases for the smallest BHs due to the smaller confinement factor $\Gamma$ under spot-limited optical pumping, as seen in Fig. \ref{fig:Thresholds_vs_theory}(a), and explained by Eq.~\eqref{eq:density_threshold}.

While this study employs optical pumping, the low internal threshold of quantum-dot-scale buried heterostructures can only be utilised with more ideal injection schemes. Importantly, electrical injection in similar nanobeam cavities has already been demonstrated \cite{jeongElectricallyDrivenNanobeam2013,crosnierHybridIndiumPhosphideonsilicon2017}. The estimated internal threshold of the lasers studied in the present work is only slightly higher than the 42~nW threshold measured at low temperature for a laser containing a single stochastically grown quantum dot~\cite{Nomura:09}. At the same time, it represents nearly an order-of-magnitude reduction compared to previously demonstrated lasers based on larger BHs~\cite{dimopoulosExperimentalDemonstrationNanolaser2023} or multiple quantum-dot layers~\cite{NOMURA20081800}.

For large carrier densities, the available gain is limited by bandfilling, an effect not fully accounted for by the logarithmic gain model\cite{morkNanostructuredSemiconductorLasers2025}. This provides a plausible explanation for why lasing was not observed for the smallest fabricated BH device ($l_{BH}=84$~nm), and indicates that improved $\Gamma$ and/or higher $Q$ will be required to reach threshold in this regime. One promising route is the use of dielectric bowtie nanocavities, which can provide substantially reduced mode volumes \cite{xiongNanolaserExtremeDielectric2025,billelauridsen2025determ}.

%Moreover, in the limit where electronic states become strongly discretised, the threshold carrier density cannot increase without bound, and is expected to remain within a small multiple of the transparency density. Consequently, further reductions in active volume can no longer be compensated by increasing carrier density \cite{morkNanostructuredSemiconductorLasers2025}. 

%\section{Discussion}
Interestingly, no significant increase in threshold is observed experimentally for larger BHs, even though these structures have a significant passivated air–BH interface area due to being etched through during cavity definition. Since surface recombination is expected to be significantly faster at these interfaces than at the buried BH–InP interfaces~\cite{berdnikovEfficientPassivationIIIAsP2025}, the observation of clear threshold behavior for all devices indicates that the surface passivation scheme can sufficiently suppress nonradiative recombination. A similar passivation scheme was previously shown to be important for achieving efficient operation in related dielectric nanolasers~\cite{xiongNanolaserExtremeDielectric2025}. While surface-related nonradiative recombination is often considered a fundamental limitation for deeply etched III--V nanostructures~\cite{bryantQuantumDotsQuantum1996}, the results demonstrate that lasing can still be achieved in this regime. These results extend buried-heterostructure nanolasers into a previously unexplored lateral size regime and demonstrate that the devices can sustain efficient emission at lateral dimensions down to 100~nm.

The present results demonstrate scaling of deterministic buried heterostructure gain media into a quantum-dot-like size regime while retaining the material quality associated with epitaxial quantum wells and the reproducibility of lithographic fabrication. Importantly, this reduction in active volume is achieved using a fabrication approach that is fully compatible with standard PhC cavity fabrication. The ability to lithographically define the position and size of a quantum-dot-scale gain region enables reproducible cavity–emitter coupling, dense integration, and systematic studies of light–matter interaction in the deep subwavelength regime \cite{billelauridsen2025determ}.

\begin{acknowledgments}
This work was supported by the Danish National Research Foundation through NanoPhoton - Center for Nanophotonics, Grant No. DNRF147
\end{acknowledgments}

\section*{Author Declarations}

\subsection*{Conflict of Interest}
The authors have no conflicts to disclose.

\subsection*{Author Contributions}

\textbf{V. Bille-Lauridsen}: Conceptualization (equal); Data Curation (lead); Formal Analysis (equal); Investigation (equal); Visualization (lead); Writing -- original draft (lead); Writing -- review and editing (equal); \textbf{R. Jarbøl}: Conceptualization (equal); Formal Analysis (equal); Investigation (equal); Writing -- review and editing (equal); \textbf{M. Xiong}: Resources (equal); Supervision (equal); Writing -- review and editing (equal); \textbf{A. Sakanas}: Conceptualization (equal); Resources (equal); Writing -- review and editing (equal); \textbf{E. Semenova}: Project Administration (equal); Resources (equal); Supervision (equal); Writing -- review and editing (equal);
\textbf{K. Yvind}: Conceptualization (equal); Project Administration (equal); Supervision (equal); Writing -- review and editing (equal);
\textbf{J. Mørk}: Conceptualization (equal); Funding Acquisition (lead); Project Administration (equal); Supervision (equal); Writing -- review and editing (equal);

\section*{DATA AVAILABILITY}
The data that support the findings of this study are openly available in DTU Data reference number 31483942. \cite{ourDATA}

\section*{References}
%\bibliography{Litterature}


\end{document}